\documentclass[twocolumn, times]{aastex631}
\usepackage{comment}

\shorttitle{AASTeX v6.3.1 Sample article}
\shortauthors{da Silva Santos et al.}

\newcommand{\uat}[2]{\href{http://astrothesaurus.org/uat/#2}{#1 (#2)}}

\DeclareRobustCommand{\ion}[2]{\textup{#1\,\textsc{\lowercase{#2}}}}
\newcommand{\asec}{$^{\prime\prime}$}

\graphicspath{{./}{figures/}}

\begin{document}

\title{Magnetic Reconnection in a Compact Magnetic Dome: Chromospheric Emissions and High-velocity Plasma Flows}

\correspondingauthor{Jo\~{a}o Santos}
\email{jdasilvasantos@nso.edu}

\author[0000-0002-3009-295X]{J. M. da Silva Santos}
\affiliation{National Solar Observatory, 3665 Discovery Drive, Boulder, CO 80303, USA}

\author[0009-0008-3051-3268]{E. Dunnington}
\affiliation{Rensselaer Polytechnic Institute, 110 8th St, Troy, NY 12180, USA}

\author[0000-0002-9309-2981]{R. Jarolim}
\affiliation{High Altitude Observatory, NSF NCAR, 3080 Center Green Dr, Boulder, CO 80301, USA}

\author[0000-0002-2344-3993]{S. Danilovic}
\affiliation{Institute for Solar Physics, Department of Astronomy, Stockholm University, AlbaNova University Centre, 106 91, Stockholm, Sweden}

\author[0000-0002-4525-9038]{S. Criscuoli}
\affiliation{National Solar Observatory, 3665 Discovery Drive, Boulder, CO 80303, USA}



\begin{abstract}

Magnetic reconnection at small spatial scales is a fundamental driver of energy release and plasma dynamics in the lower solar atmosphere. We present observations of a brightening in an active region, captured in high-resolution data from the Daniel K. Inouye Solar Telescope (DKIST) using the Visible Broadband Imager (VBI) and the Visible Spectro-Polarimeter (ViSP). The event exhibits Ellerman bomb-like morphology in the H$\beta$ filter, associated with flux cancellation between a small negative polarity patch adjacent to opposite-polarity plage. Additionally, it displays enhanced emissions in \ion{Ca}{II} K, hot elongated features containing Alfvénic plasma flows, and cooler blue-shifted structures.
We employ multi-line, non-local thermodynamic equilibrium (non-LTE) inversions of the spectropolarimetric data to infer the stratification of the physical parameters of the atmosphere. Furthermore, we use the photospheric vector magnetogram inferred from the ViSP spectra as a boundary condition for nonlinear force-free field extrapolations, revealing the three-dimensional distribution of squashing factors. We find significant enhancements in temperature, velocity, and microturbulence confined to the upper photosphere and low chromosphere. Our findings provide observational evidence of low-altitude magnetic reconnection along quasi-separatrix layers in a compact fan-spine-type configuration, highlighting the complex interplay between magnetic topology, energy release, and plasma flows. 
\end{abstract}



\section{Introduction} 
\label{sec:intro}

Small-scale ($\lesssim$\,1\asec) magnetic reconnection events are fundamental and dynamic processes shaping the lower solar atmosphere. These events, which often result from complex interactions of opposite magnetic polarities, are observed as transient (lasting from a few seconds to a few minutes), dot-like or elongated brightenings across multiple spectral bands from ultraviolet (UV) and visible wavelengths \citep[e.g.,][]{2002ApJ...575..506G,2014Sci...346C.315P,2015ApJ...812...11V,2019ApJ...887...56T} to the millimeter continuum \citep[][]{2022A&A...661A..59D}. Despite their small scale, these phenomena have implications for understanding larger solar processes, such as solar eruptions, while offering a unique opportunity to investigate the physics of magnetic reconnection.

Ellerman bombs \citep[EBs;][]{1917ApJ....46..298E} serve as prototypical examples of magnetic reconnection in the photosphere. Observationally, they are characterized by enhanced, flickering emissions in the wings of hydrogen Balmer and \ion{Ca}{II} lines, and they are usually located between or near magnetic bipoles \citep[][and references therein]{2013JPhCS.440a2007R}. 
UV bursts (UVBs) belong to an analogous broad class of events, notably characterized by brightenings displaying broad line profiles in transition region temperature lines. These profiles exhibit velocities of a few hundred $\rm km\,s^{-1}$, comparable to the Alfvén speed, while the chromospheric lines are also enhanced \citep[][and references therein]{2018SSRv..214..120Y}. 
Numerical and observational studies suggest that EBs and UVBs represent different manifestations of a shared underlying process, with variations in the height and physical conditions of the reconnection site, as well as the viewing angle, determining their spectral signatures and co-occurrence \citep[e.g.,][]{2017ApJ...839...22H,2019A&A...626A..33H,2019ApJ...875L..30C,2020A&A...633A..58O,2023A&A...672A..47S}. 

Key open questions in the study of EBs/UVBs center on understanding the magnetic topology and nature of the reconnection processes, as well as explaining the dynamics of the observed outflows. For example, the tearing mode instability, or plasmoid-mediated reconnection, in which the current sheet fragments into smaller magnetic islands \citep[reviewed in ][]{2022LRSP...19....1P}, has been proposed to explain multi-component velocities in spectroscopic observations \citep[][]{2015ApJ...813...86I} and the intermittency of UV emissions \citep[][]{2019A&A...628A...8P}. High-resolution ground-based observations have occasionally resolved the sources of this intermittency into fine-scale ($\sim$0.1\asec) recurrent plasmoid-like bright blobs, which appear to be ejected from reconnection sites \citep[][]{2017ApJ...851L...6R,2023A&A...673A..11R,2021A&A...647A.188D}. 
Alternatively, turbulent reconnection driven by the Kelvin-Helmholtz instability has been suggested as a more plausible mechanism in certain contexts, given the large apparent size of the reconnection region in some EB events \citep[][]{2024NatCo..15.8811A}. Enhanced turbulence has also been invoked to explain excess line widths attributed to turbulent motions within the reconnection region in UV brightenings \citep[e.g.,][]{2019ApJ...885..158W,2020ApJ...890L...2C,2020A&A...633A..58O}. 

With regard to the magnetic topology, U-type (concave-up) configurations, where the reconnection occurs in bald patch separatrices \citep[][]{1993A&A...276..564T}, are typically consistent with observations of EBs \citep[e.g.,][]{2002ApJ...575..506G,2004ApJ...614.1099P,2009ApJ...701.1911P}. The same configuration has been proposed to explain some UVBs \citep[][]{2014Sci...346C.315P,2017ApJ...836...52Z}. However, in some instances, UVBs appeared to be triggered at magnetic null points at the top of magnetic domes in fan-spine configurations \citep{2017A&A...605A..49C}. This configuration has also been observed in larger-scale flares exhibiting circular flare ribbons \citep[e.g.,][]{2009ApJ...700..559M,2012ApJ...760..101W,2013ApJ...778..139S}, which suggests a link between the weakest and strongest reconnection events \citep[][]{2018A&A...617A.128S}. Recently, a study of a large sample of quiet-Sun EBs and associated UV brightenings reported a range of magnetic configurations, spanning from simple bipoles to fan-spine topologies, where the EB and UVB components originate from different parts of the 3D magnetic structures, as inferred from potential field extrapolations \citep{2024arXiv241203211B}. 

In this study, we present the analysis of a small-scale brightening observed in an active region (AR) with the Daniel K. Inouye Solar Telescope \citep[DKIST;][]{2020SoPh..295..172R}. We derive the atmospheric parameters of the event through multi-line non-local thermodynamic equilibrium (NLTE) inversions and perform non-linear force-free field (NLFFF) extrapolations based on a high-resolution ViSP magnetogram, complementing the previous topological studies and providing new observational insights into the impact of magnetic reconnection in the low atmosphere.

\section{Data}
\label{sec:Observations}

\subsection{Data Acquisition and Post-processing}

We used imagery provided by the Visible Broadband Imager \citep[VBI;][]{2021SoPh..296..145W} and spectropolarimetric data acquired by the Visible Spectro-Polarimeter \citep[ViSP;][]{2022SoPh..297...22D} instruments at the DKIST. The target was AR NOAA 13465 close to the disk center at $\mu\sim$\,0.99, where $\mu$ is the cosine of the heliocentric angle, on 2023 October 16. The observing campaign (proposal 2\_114) consisted of multiple ViSP scans of different sub-fields around the AR both in spectroscopy and polarimetry modes supported by VBI imaging \citep[][]{2025arXiv250202742S}. Here, we focused on a particular pointing centered on the sunspot at helioprojective coordinates $\sim$\,[-76\asec, 90\asec] between 20:18\,$-$20:38\,UT in polarimetry mode.   
We also used line-of-sight (LOS) magnetograms provided by the Helioseismic and Magnetic Imager \citep[HMI;][]{2012SoPh..275..207S} and EUV and UV images taken by the Atmospheric Imaging Assembly \citep[AIA;][]{2012SoPh..275...17L} on board the Solar Dynamics Observatory \citep[SDO;][]{2012SoPh..275....3P}.

The VBI data comprise high-resolution broad-band images in the G-band (4305\,\AA), H$\beta$ (4861\,\AA), and \ion{Ca}{II} K (3933\,\AA) filters. The VBI acquired sequences of 80 images with single exposure times of 0.7\,ms, 3.5\,ms, and 25\,ms for the G-band, H$\beta$, and \ion{Ca}{II} K filters, respectively. The VBI operated at a higher cadence of 2.7\,s in the G-Band filter, interspersed with a lower cadence sub-cycle in the other filters every fifth image. The cadence for \ion{Ca}{II} K and H$\beta$ was 19.8\,seconds. The image scale is approximately 0.0106\asec~per pixel. The filters' full width at half maxima (FWHM) are approximately 4.37\,\AA, 0.46\,\AA, and 1.01\,\AA, for G-band, H$\beta$, and \ion{Ca}{II} K filters, respectively.
As part of the reduction pipeline, the images were dark- and gain-corrected and processed using a speckle algorithm to mitigate seeing-induced distortions \citep[][]{2008A&A...488..375W}. However, the quality of the reconstructed timeseries varies with the seeing conditions. 
We coaligned the VBI and SDO data via cross-correlation between the VBI G-band images and the HMI 6173\,\AA~continuum images, and we coaligned VBI and ViSP by cross-correlating the G-band images and the 6301\,\AA~continuum images from the ViSP. Additionally, we cross-correlated the VBI channels among themselves to corrected for measured relative misalignments on the order of $\sim$\,0.1\asec$-$\,0.2\asec, consistent with wavelength-dependent atmospheric dispersion \citep[][]{2006SoPh..239..503R}.

The ViSP data consist of a single spectropolarimetric raster scan in three passbands, approximately 13–18\,\AA~wide, centered on the \ion{Fe}{I} 6301 and 6302\,\AA~lines (hereafter $\lambda6301(2)$), the \ion{Na}{I} $\rm D_{1}$ 5896\,\AA~line (hereafter $\lambda5896$), and the \ion{Ca}{II} 8542\AA~line (hereafter $\lambda8542$). The raster maps covered an area of about 46\asec$\times$76\asec~in $\lambda6301(2)$, 46\asec$\times$61\asec~in $\lambda5896$, and 46\asec$\times$49\asec~in $\lambda8542$ encompassing the sunspot. The raster step was $\sim$\,0.1\asec. The pixel scale along the slit varies between the three arms; for analysis purposes, we rebinned the spectra (factors of $\sim$\,3$-$5) to 0.1\asec~square pixels. Then, we co-aligned the three arms by cross-correlation of continuum images from each arm. The number of modulation states and cycles was 10, requiring $\sim$\,3\,s per slit position, resulting in a total raster time of approximately 20 minutes. We estimated the polarimetric sensitivity to be on the order of $(6-8)\times10^{-4}$ relative to the continuum, with better signal-to-noise ratio in the photospheric passbands. 

The ViSP data were reduced with version 2.10.7\footnote{\url{https://docs.dkist.nso.edu/projects/visp/en/v2.10.7/}} of the calibration pipeline, which includes dark and gain corrections, as well as polarization calibration. We further corrected for residual polarization crosstalk using standard methods \citep{1992ApJ...398..359S}. The wavelength and absolute intensity calibrations were performed via cross-correlation of the average spectra with the solar atlas of \citet{1984SoPh...90..205N}. This included an optimization procedure to improve the dispersion values upon what is provided in the Level 1 file headers using the wavelength positions of the telluric lines as a reference. We obtained approximately 0.014\,\AA, 0.013\,\AA, and 0.019\,\AA~for the $\lambda5896$, $\lambda6301(2)$, and $\lambda8542$ passbands, respectively. 

The level 1 AIA data were converted into level 1.5 with the \texttt{aia\_prep} routine in SolarSoftware \citep[][]{1998SoPh..182..497F}. We show AIA 1700\,\AA~and 304\,\AA~images taken at a cadence of 24\,s  and 12\,s, respectively. The image scale is 0.6\asec per pixel. The HMI LOS magnetograms have a cadence of 45\,s and image scale of 0.5\asec per pixel. We also used one 720\,s-cadence vector magnetogram taken at 20:34\,UT that was processed with the SHARP pipeline \citep[][]{2014SoPh..289.3549B}.

\subsection{Target Overview}

\begin{figure*}[t]
    \centering
    \includegraphics[width=\linewidth]{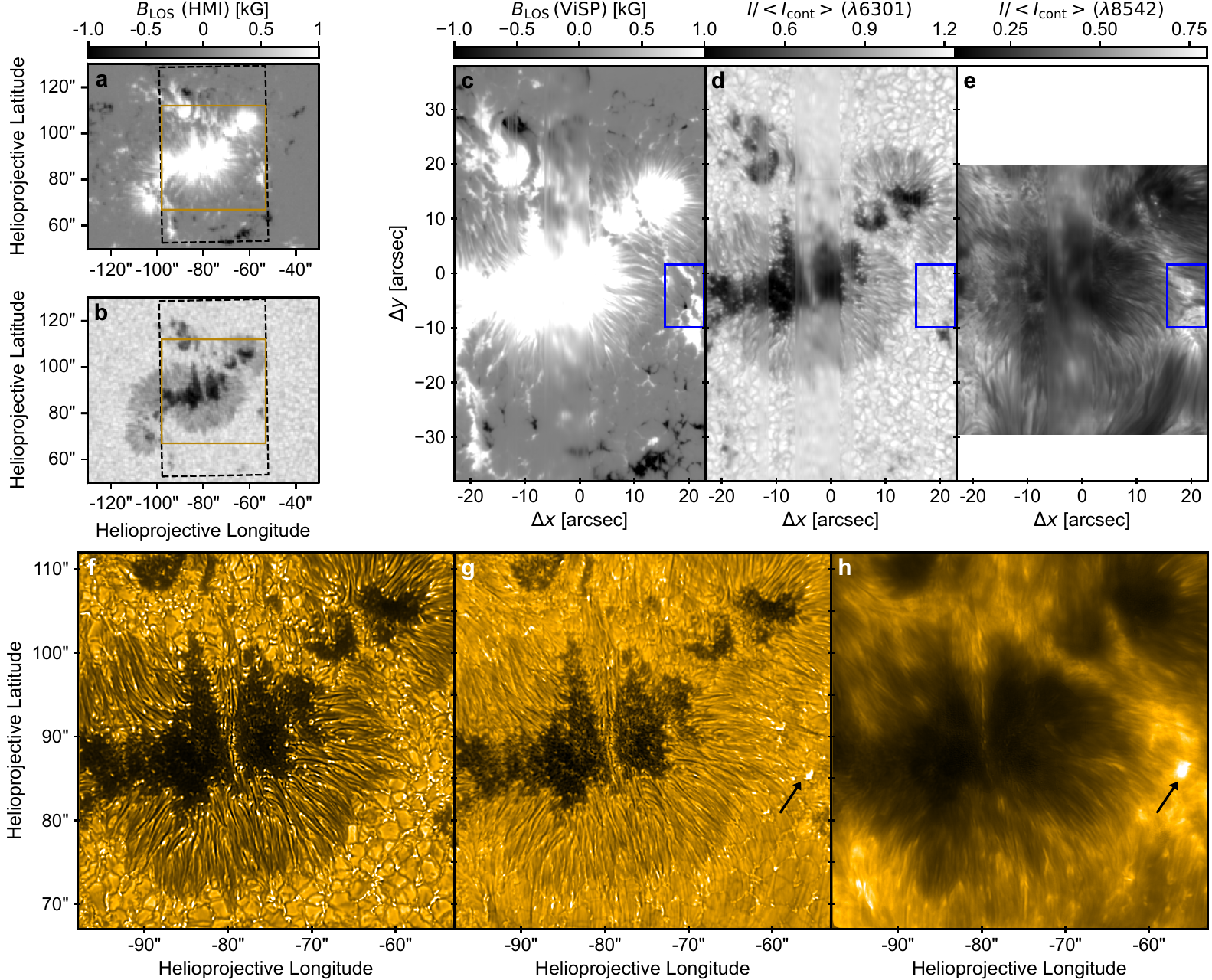}
    \caption{Overview of AR 13465 on 2023 October 16. (a): SDO/HMI LOS magnetogram. (b): SDO/HMI 6173\,\AA~continuum. (c): DKIST/ViSP photospheric LOS magnetogram obtained with \texttt{PyMilne}. (d): DKIST/ViSP raster in the 6301\AA~continuum. (e): DKIST/ViSP \ion{Ca}{II}\,8542\,\AA~core. (f): DKIST/VBI G-band. Panel (g): DKIST/VBI H$\beta$. (h): DKIST/VBI \ion{Ca}{II}\,K. All colormap ranges are capped for display purposes. The dashed and solid boxes in (a)--(b) show the ViSP ($\lambda6301(2)$) and VBI fields of view, respectively. The blue boxes in panels (c)--(e) show the region selected for the NLTE inversions (Fig.\,\ref{fig:inv}). The arrows indicate the brightening that is focus of study here.} 
    \label{fig:overview}
\end{figure*}

Figure~\ref{fig:overview} presents an overview of the target. Panels (a) and (b) show SDO/HMI observations, providing a contextual view of the AR, which is only partially captured by DKIST. The magnetic flux density distribution in the field of view (FOV) is dominated by the sunspot's positive polarity, surrounded by plage regions and pores of opposite polarity.
DKIST/ViSP observations zoom into the region enclosed by the dashed box in (a). Panels (c) and (d) show the ViSP LOS magnetogram obtained using Milne-Eddington inversions of the $\lambda6301(2)$ lines (described in Section\,\ref{sec:inversions}), and the 6301\,\AA~continuum intensities for direct comparison with (a) and (b), showing a good agreement. While the raster map is affected by periods of poor seeing, the subregion within the blue box -- the primary focus of this paper, clearly reveals a small negative polarity patch ($\lesssim$\,1\asec) near a pore of positive polarity. This negative patch is barely distinguishable in the HMI magnetogram. Panel (e) shows the intensity map in the core of $\lambda8542$, revealing the opaque chromospheric fibrils. 

Panels (f)–(h) present the images from DKIST’s VBI taken at a specific time within the ViSP raster interval. The G-band image shows very fine details of the penumbrae and the magnetic elements associated with the bright points. The $\sim$0.46\,\AA~wide H$\beta$ filter captures both the photosphere and chromosphere, revealing fibril structures extending outward from both penumbrae and plage regions. Finally, the \ion{Ca}{II}\,K images also provide a combined view of the photosphere and chromosphere but with higher chromospheric opacity than H$\beta$. The arrows in these panels mark the brightenings analyzed in this study, which occur at the interface between opposite magnetic polarities. The blue box in panels (c)–(e) defines the region of interest (ROI).

\section{Methods}
\label{sec:Methods}

\subsection{Spectropolarimetric Inversions}

To obtain the magnetic field at different heights, we began by applying two different inversion techniques to the spectropolarimetric data, focusing on the main spectral diagnostics within the three ViSP arms. We employed the Milne-Eddington (ME) approximation to the $\lambda$6301(2) lines using the \texttt{pyMilne} code \citep{2019A&A...631A.153D}, and we applied the Weak Field Approximation (WFA) to the cores of $\lambda$5896 and $\lambda$8542 lines using the \texttt{spatial\_WFA} code \citep{2020A&A...642A.210M}. 
The resulting magnetic field estimates can then serve as inputs for more detailed NLTE inversions (see below) or magnetic field extrapolations (Section~\ref{sec:extrapolation}).

The inversions of $\lambda$6301(2) successfully retrieved the magnetic field components in the photosphere; the LOS magnetogram derived from $\lambda$6301(2) is displayed in Fig.\,\ref{fig:overview}c. We further resolved the magnetic azimuth angle ambiguity using the minimum energy method with the \texttt{AMBIG} code \citep[][]{2014ascl.soft04007L}. We verified that the resulting magnetic vector components were generally consistent with the cotemporal, lower resolution SHARP maps. The WFA inversions, however, lacked consistency in parts of the sunspot umbra, likely due to the high field strengths ($\gtrsim$\,2\,kG). As this could negatively impact the extrapolations, we excluded the $\lambda$5896 and $\lambda$8542 WFA inversion maps from subsequent analysis. However, we still include those lines in the NLTE inversions of the ROI, as the field strength is not a constraint for this method.

We performed NLTE inversions using the STockholm inversion Code\footnote{\url{https://github.com/jaimedelacruz/stic}} \citep[\texttt{STiC};][]{2016ApJ...830L..30D,2019A&A...623A..74D}, which is a massively parallel, multi-atom inversion code based on a modified version of the Rybicki \& Hummer \citep[\texttt{RH};][]{2001ApJ...557..389U} radiative transfer code. The atomic models used in the statistical equilibrium calculations included a 11-level atom for \ion{Na}{I} plus a \ion{Na}{II} continuum level, and a 5-level model for \ion{Ca}{II} plus a \ion{Ca}{III} continuum level. The \ion{Fe}{I} 6301 and 6302\,\AA~lines were treated in LTE. For expediency, hydrogen was treated in LTE, which is a fair assumption in the layers probed by the $\lambda5896$ and $\lambda8542$ lines from an inversion standpoint \citep[][]{2024arXiv240815908D}. 

We considered incorporating additional weak photospheric lines of \ion{Fe}{I} (5892.7, 5898.2, 8538\,\AA), \ion{Si}{I} (8536.1\,\AA), \ion{Ni}{I} (5892.9\,\AA), and \ion{Ti}{I} (8548.1\,\AA) within the ViSP passbands using atomic data from the Kurucz database\footnote{\url{http://kurucz.harvard.edu/}}. 
However, inversion tests with the AR data showed systematic misfits in intensity (and sometimes wavelength) to these lines to varying degrees, suggesting atomic data inaccuracies. Therefore, we excluded these lines from the inversions, fitting only the strongest lines in each ViSP arm. Nonetheless, we included them in spectral synthesis to assess how well they are predicted by the inferred models.

The inversion process followed a standard multi-cycle approach, progressively increasing the number of nodes to improve the fit. Here, we ran three cycles with eight nodes for temperature, increasing up to six nodes for line-of-sight velocity and microturbulence, and up to three nodes for the line-of-sight and transverse components of the magnetic field, as well as the azimuth angle. Increasing the number of nodes beyond this did not significantly improve the quality of the fits. We employed the ME inversion maps mentioned above as an initial guess for the magnetic field stratification and we used the F99 model \citep[][]{1999ApJ...518..480F} as an initial guess for the temperature stratification. F99 reproduces the (average) QS atlas profiles of $\lambda5896$ and $\lambda8542$ quite well using the aforementioned atomic models. We present the synthesis of these lines from F99 alongside the brightening line profiles to highlight their differences from QS.

\subsection{Magnetic Field Extrapolations}
\label{sec:extrapolation}

We performed NLFFF extrapolations based on photospheric vector magnetograms to estimate the 3D topology of the AR. To this aim, we applied the method\footnote{\url{https://github.com/RobertJaro/NF2}} developed by \cite{jarolim2023nf2}, which employs a Physics Informed Neural Network \citep[PINN; ][]{raissi2019pinns} to estimate the 3D magnetic field, $\vec{B}$. An iterative optimization scheme was used to satisfy the boundary conditions and enforce the force-free equations ($\nabla \cdot \vec{B}=0$; $\nabla \times \vec{B} \times \vec{B}=0$). This approach has shown the ability to flexibly find a trade-off between noisy observational data and the incomplete physical model, providing realistic approximations of the coronal magnetic field \citep[cf.,][]{Purkhart2023m4_flare, McKevitt2024nonthermal, korsos2024preeruptive, jarolim2024ar13664}. We made use of the mesh-free simulation volume which allows us to flexibly embed a high-resolution DKIST/ViSP vector magnetogram as part of the AR (inferred from the $\lambda$6301(2) observations) into the lower resolution, but more extended, SDO/HMI vector magnetogram (i.e., the SHARP maps in cylindrical equal area projection). This is essential for accurately modeling the field connectivity within the global topology of the AR. Therefore, we can estimate the small scale magnetic field structure of the ViSP observation in the ROI, while simultaneously obtaining the global magnetic topology of the AR magnetic field. 
Per updating step, we randomly sampled $2^{13}$ points from both the SHARP map at 360\,km per pixel resolution and the DKIST/ViSP observation at 72\,km per pixel resolution. Here, we masked out the ViSP FOV from the SHARP map. Similarly to \cite{jarolim2023nf2}, we continuously sampled $2^{14}$ random points from the simulation volume. Per updating step, we optimized both for the sampled boundary conditions and minimized the deviation from the partial differential equations $\nabla \cdot \vec{B}=0$, $\nabla \times \vec{B} \times \vec{B}=0$, where we weight the SHARP boundary condition with $1$ and the divergence and force-free equations with $10^{-2}$. For the ViSP boundary condition, we set the weighting factor to $1$ and exponentially increase the weight to $10$ over $5\times10^4$ iterations. The optimization was performed over $2\times10^5$ iterations until convergence was reached. 

The magnetic field solution can be sampled at arbitrary resolution from the resulting neural representation. For further evaluation, we sampled a frame spanning $30\times30$\,Mm$^2$ around the negative polarity, with a resolution of 72\,km per pixel. We computed the current density, $\vec{J} = (c/4 \pi) \nabla \times \vec{B}$ (where $c$ is the speed of light), using smooth derivatives from the neural representation. Additionally, we computed the squashing factor, $Q$, and the twist number, $\mathcal{T}_w$, for each point in the resulting 3D volume using the \texttt{FastQSL}\footnote{\url{https://github.com/peijin94/FastQSL}} method \citep{Zhang2022fastqsl}. The squashing factor is used to characterize the magnetic topology, where a large $Q$ indicates that local regions experience a strong divergence of magnetic field lines \citep[e.g.,][]{Titov2007}. The twist number is a measure of how many turns two infinitesimally close magnetic field lines make around each other \citep[][]{2006JPhA...39.8321B} and is used to identify variations in helicity.

\begin{figure*}
    \centering
    \includegraphics[width=\linewidth]{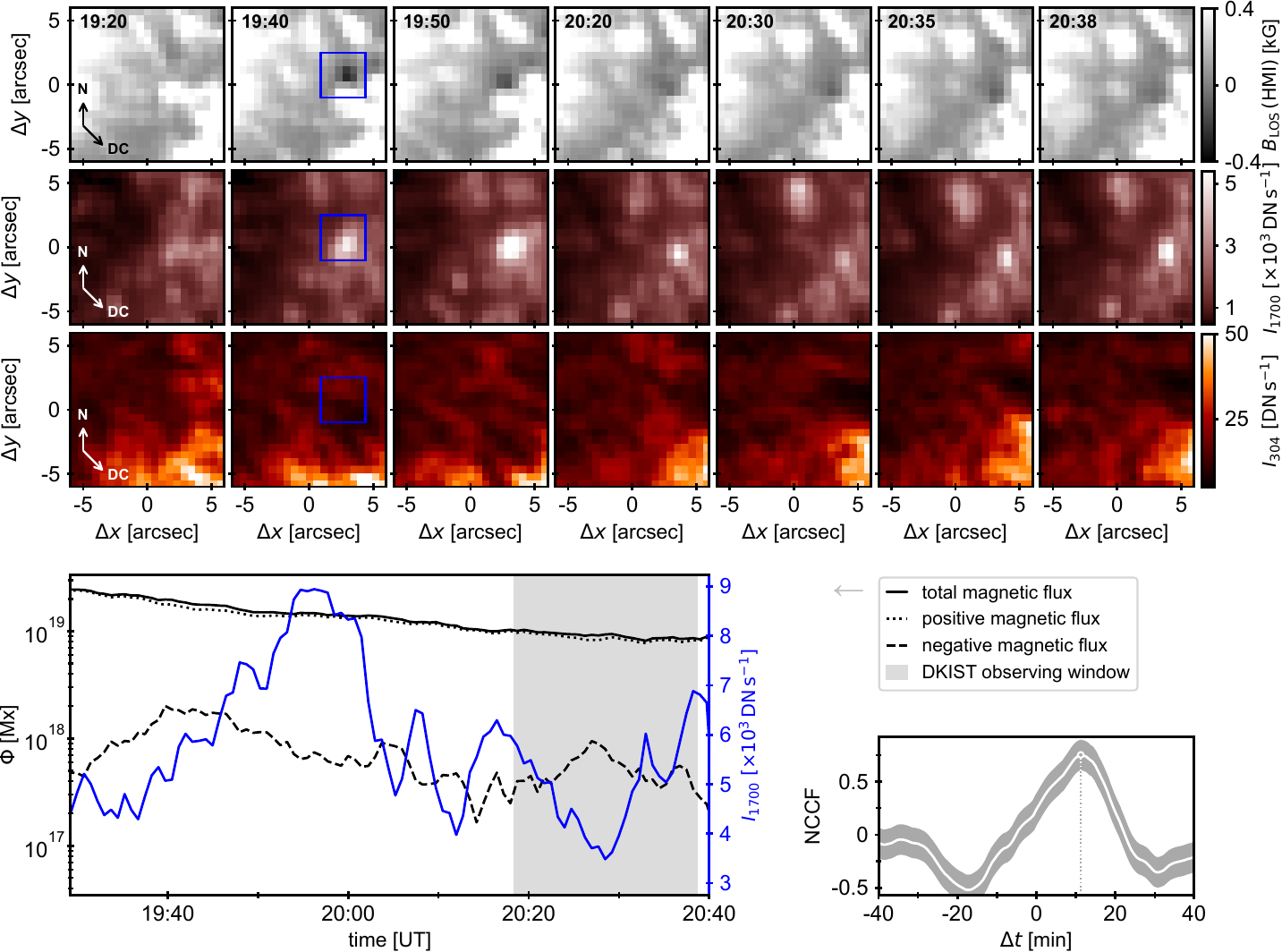}
    \caption{Time evolution of the brightening captured by SDO. The upper panels show, from the top to the bottom, HMI line of sight magnetograms capped at $\pm$\,0.4\,kG, and intensities in the AIA 1700\,\AA~and 304\,\AA~ channels of the brightening highlighted in the bottom panels in Fig.~\ref{fig:overview}; the arrows indicate the directions of solar north and disk center. Lower left panel: time variation of total, positive, and negative magnetic fluxes and maximum AIA 1700\,\AA~intensities within the region enclosed by the blue box in the top panels. Lower right panel: normalized cross-correlation function between AIA and HMI negative flux time-series, with 95\% confidence region represented by the shaded area.} 
    \label{fig:timeseries}
\end{figure*}

\begin{figure*}[t]
    \centering
    \includegraphics[width=\linewidth]{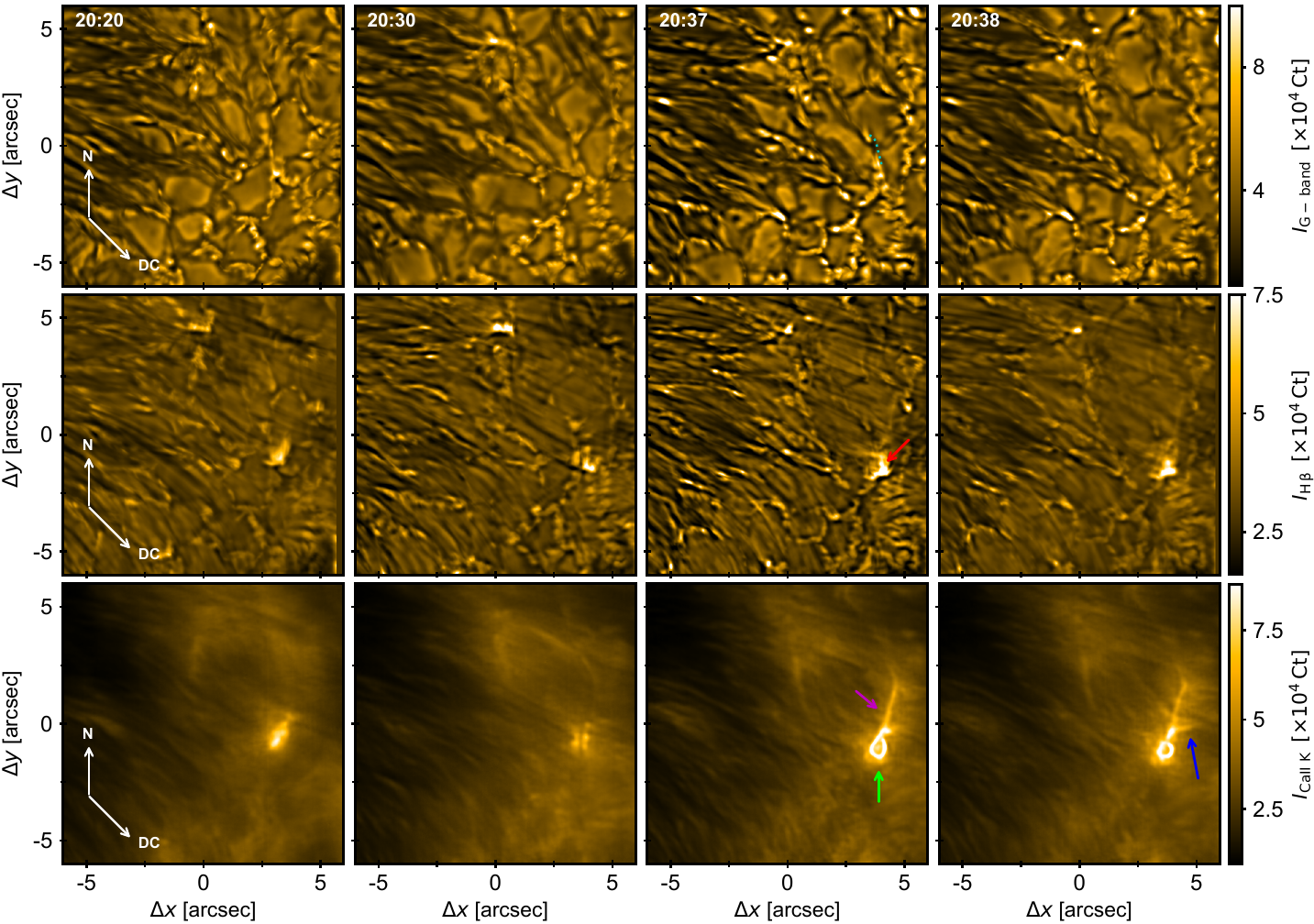}
    \caption{High-resolution view of the event provided by DKIST/VBI. From the top to the bottom, intensities (in counts) in the VBI G-band, H$\beta$ and \ion{Ca}{II} K filters. The FOV is the same as in Fig.\,\ref{fig:timeseries}. The white arrows indicate the directions of solar north and disk center. The colored arrows point to different features of interest. The bright annular structure in \ion{Ca}{II} K from 20:37\,UT is an image reconstruction artifact. A three-column animated version of this figure is available, where the cyan dotted lines show \ion{Ca}{II}\,K intensity contours.} 
    \label{fig:timeseries2}
\end{figure*}

\section{Results}

\subsection{Observational Characteristics}
\label{sec:obs_stats}

Figure~\ref{fig:timeseries} presents selected time frames of the brightening observed by AIA, along with the magnetic flux variation from HMI magnetograms over a longer time span than DKIST observations. The image sequences were corrected for solar rotation. The magnetic flux, $\Phi$, was calculated within a 3.5\asec$\times$3.5\asec~region centered on the negative polarity patch concentration at 19:40\,UT (blue box).
The HMI magnetograms reveal the emergence of a small-scale negative polarity patch, moving approximately radially outward from the sunspot penumbra in the southeast direction. This motion is characteristic of moving magnetic features (MMFs). The patch enters a predominantly positive polarity region, with the negative magnetic flux reaching its maximum around 19:40 UT.
Using centroid tracking and an uncertainty of half the AIA image scale, we measured its plane-of-sky velocity as 1.8($\pm$0.3) $\rm km\,  s^{-1}$, consistent with typical MMF speeds \citep[e.g.,][]{2005ApJ...635..659H,2019ApJ...876..129L}.

By the time DKIST began observing $\sim$\,20\,min later, the negative polarity patch had become almost indistinguishable (top panels), though it remains visible in the ViSP magnetogram (cf. Fig.~\ref{fig:overview}c). Still, significant variations in the magnetic flux could be measured from HMI (bottom panel in Fig.~\ref{fig:timeseries}). The total magnetic flux within the blue box decreased steadily over the course of an hour at an average rate of $d\Phi/dt$\,$\sim$\,10$^{19}\rm\,Mx\,h^{-1}$, that is $\sim$\,$3\times$10$^{15}\rm\,Mx\,s^{-1}$, comparable to reported flux cancellation rates in ARs \citep[e.g.,][]{2016ApJ...823..110R,2018ApJ...854..174T,2020A&A...644A.130C}.

The appearance of the opposite polarity was followed by a significant increase in the 1700\,\AA~continuum emission between the two polarities, albeit with a temporal offset. Through cross-correlation, we determined a lag of approximately 11.3\,min between the peak AIA intensities and the negative magnetic flux, with a correlation coefficient of 0.75. This delay may be attributed to the time required for the build up of opposite polarity flux to induce localized heating through magnetic reconnection with the ambient field. This timescale is consistent with 3D MHD simulations, which demonstrate that rising magnetic loops take several minutes to interact and form reconnecting current sheets before energy release occurs \citep[][]{2017A&A...601A.122D,2019A&A...626A..33H} and that would be detected in the form of enhanced UV (and other) emissions. \citet[][]{2024A&A...686A.218N} recently reported a similar 11-min-lag between the appearance of a photospheric magnetic bipole and EB brightenings.

We find no significant enhancements in the hotter AIA channels (e.g, 171\,\AA, 94\,\AA) at the same location. However, we do observe a slight dimming in AIA 304\,\AA~(bottom panels), which could be due to absorbing cool material ejected into the upper atmosphere. 

\subsubsection{Photospheric Dynamics in High Resolution}

Figure~\ref{fig:timeseries2} displays selected time frames of the VBI time series data in the same FOV as in Figure~\ref{fig:timeseries}. We also provide a supplementary movie showing the time evolution in greater detail. The VBI images resolve finer details of the brightenings compared to AIA, showing EB-like "flame" structures in the H$\beta$ filter \citep[e.g.,][]{2017A&A...598A..33L,2020A&A...641L...5J} in two different events within the top right quadrant of the FOV (e.g., 20:30\,UT). The EB classification is corroborated by enhanced AIA 1700\,\AA~emissions at the same locations \citep[e.g.,][]{2013ApJ...774...32V,2019A&A...626A...4V, 2017ApJS..229....5D}, as shown in Fig.~\ref{fig:timeseries}. The northernmost EB candidate is short-lived, lacks a \ion{Ca}{II} K counterpart, and does not appear to be linked to a photospheric magnetic bipole. We therefore focus on the southernmost event, which corresponds to the AIA brightening within the blue box in Fig.\,\ref{fig:timeseries} and displays more distinct characteristics.
The evolution of substructures in the ROI (red arrow) resembles small-scale ($\sim$\,0.1\asec$-$0.4\asec) moving blobs, interpreted as plasmoids in rare high-resolution observations of similar events in the H$\alpha$ line \citep[][]{2023A&A...673A..11R}, the \ion{Ca}{II} K line \citep[][]{2017ApJ...851L...6R,2021A&A...647A.188D}, and more recently in the \ion{He}{I} 10830\,\AA~line \citep[][]{2025arXiv250110246L}. However, direct comparison with the literature is not trivial because differences in spectral sampling (broad-band vs narrow-band images) could affect the visibility and morphology of fine-scale features. Additionally, the limited cadence of our observations imposes constraints on tracking the evolution and motion of these structures.

G-band images reveal fine "filigree" \citep[][]{1973SoPh...33..281D} structures at the location of the southernmost EB candidate. As bright filigree converge at the apex of several granules and the granules evolve, the flame-like feature in H$\beta$ changes morphology. This phenomenon has been reproduced by high-resolution simulations, showing that flames outline current sheets forming where complex magnetic structures are compressed \citep{2017A&A...601A.122D,2017ApJS..229....5D}. The G-band movie shows that around 20:28\,UT, granules extend northwest from the H$\beta$ structure. Bright magnetic elements rapidly approach from the northwest (parallel to the cyan dotted line, Fig.~\ref{fig:timeseries2} 20:37\,UT), with plane-of-sky velocities of approximately $\sim$\,4$-$6\,$\rm km\,s^{-1}$. This indicates that the new magnetic flux is advected towards the same location. When one of the bright G-band points merges with the filigree at the site of the H$\beta$ enhancement, a distinct circular structure appears in the \ion{Ca}{II} K filter. 

\subsubsection{Chromospheric Brightenings}

The brightening in the \ion{Ca}{II} K filter initially appears as a diffuse blob, spatially offset from the H$\beta$ enhancement in the limb-ward direction. Later, a bright ring (green arrow) appears connected to a bright fibril-like structure (magenta arrow) oriented orthogonally to the dark fibril background. The ring structure was found to be an artifact of the speckle image reconstruction\footnote{\url{https://nso.atlassian.net/wiki/spaces/DDCHD/pages/3480518672/VBI+Data+Set+Caveats}}, as the raw images are saturated at the very center of the brightening. 
We opted to show the \ion{Ca}{II} K images in Fig.~\ref{fig:timeseries2}, as the bright annulus offers a proxy for the size and shape of the brightening in that filter. The rest of the image features are reliable.
Furthermore, the VBI observations also reveal jet-like structures with projected widths of $\sim$\,0.2\asec~(blue arrow) forming at an angle to the bright fibril-like structure. However, these features may not be jets in a strict sense, but a manifestation of the interaction between the EB and overlying canopy visible in the \ion{Ca}{II} K. This interpretation is uncertain given the quality of the image sequence.
The bright fibril is also visible in the H$\beta$ images but appears much dimmer; for example, compare 20:30 and 20:37\,UT.

The configuration of the G-band bright points roughly mirrors the temporal evolution of the \ion{Ca}{II} K ring, albeit with a spatial offset in the limbward direction likely due to projection effects. This suggests that the photospheric footpoints associated with the chromospheric emissions may be confined to a region of only $\sim$\,1\asec. 
The continuous inflow of magnetic bright points throughout the evolution of the \ion{Ca}{II} K structures may account for the deposited energy, as it introduces magnetic flux at high horizontal velocity. We further explore the magnetic topology of the event in Section\,\ref{sec:extrapol}. 

\begin{figure*}[t]
    \centering
    \includegraphics[width=\linewidth]{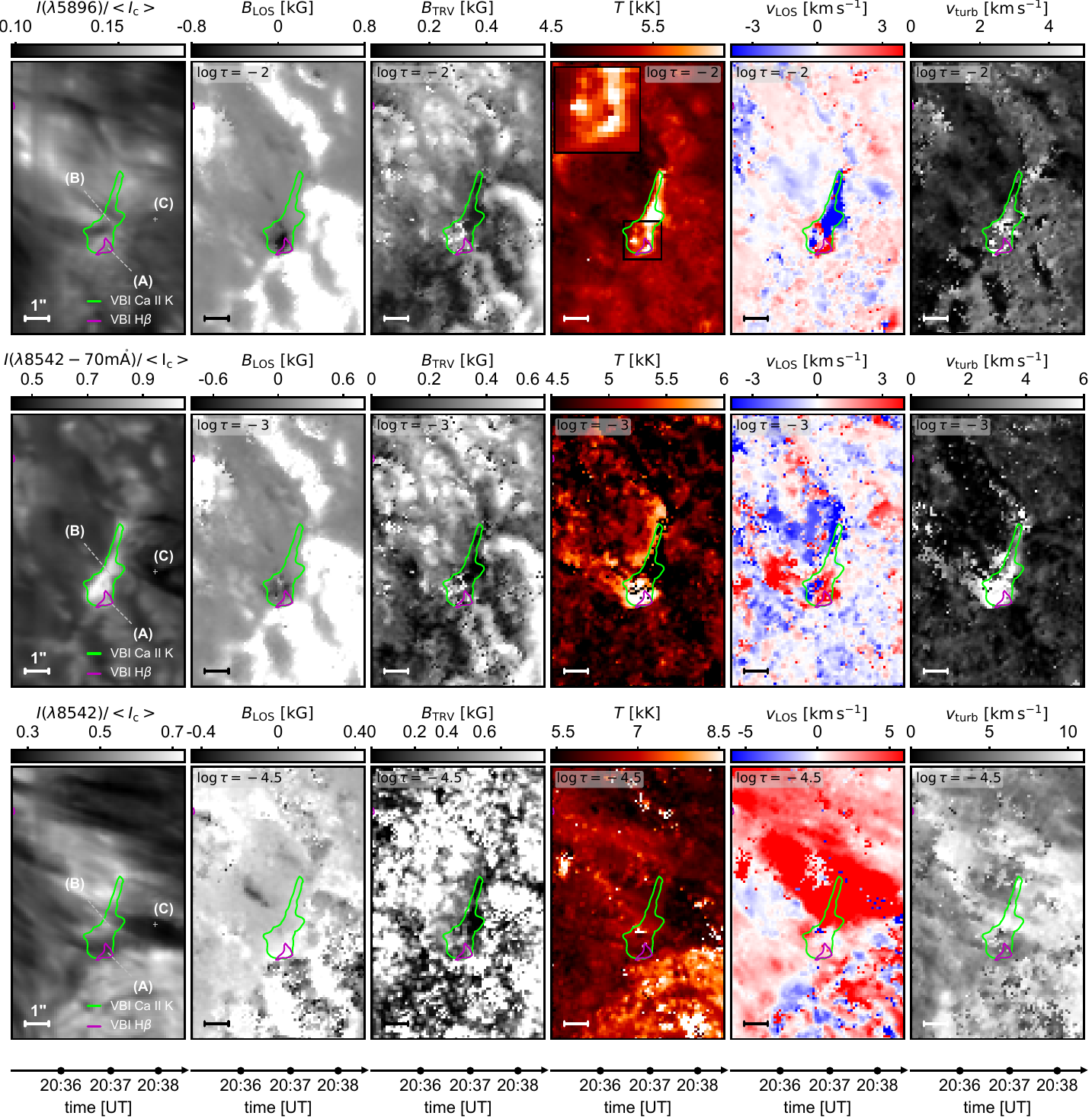}
    \caption{Atmospheric stratification from NLTE inversions. The leftmost panels display the intensities in the $\lambda$5896 core, as well as core and blue wing of $\lambda$8542 observed by the ViSP within the blue box shown in Fig.\,\ref{fig:overview}(c--e). The green(magenta) contours outline the brightness enhancements in the VBI \ion{Ca}{II}\,K (H$\beta$) images at 20:37\,UT (cf. Fig.~\ref{fig:timeseries2}). The other panels, from left to right: LOS and transverse components of the magnetic field, temperature (with an inset zoom of the central region), LOS velocity, and microturbulence at three different optical depths. The y-axes are aligned with solar north. The colormap ranges are capped for display purposes. The timestamps corresponding to particular slit positions are shown at the bottom.} 
    \label{fig:inv}
\end{figure*}

\subsection{NLTE Inversions}
\label{sec:inversions}

Figure~\ref{fig:inv} presents the results of the NLTE inversions of the ROI at three different optical depths. We note that we cannot infer the atmosphere parameters beyond $\log\tau$\,$\sim$\,$-4.5$ from these observations, as the line response functions are weak at lower optical depths. The inversion maps were filtered with a Wiener filter to mitigate inversion noise. For context, we also show the ViSP intensities in the $\lambda5896$ and $\lambda8542$ line cores and blue wing of $\lambda8542$ (leftmost panels), as well as VBI intensity contours outlining the shape of the brightenings (at about half-maximum level) in \ion{Ca}{II} K (green) and H$\beta$ (magenta) at 20:37\,UT (cf. Fig.~\ref{fig:timeseries2}) when the ViSP slit was scanning the region (see ViSP arrow of time at the bottom). In the discussion below, we also refer to the complementary Fig.~\ref{fig:fits} showing the inversion results in greater detail for three selected locations: (A) at the center of the H$\beta$ brightening, (B) at the upper part of the brightening, and (C) at an absorbing feature in $\lambda$8542. 
While the morphology of the VBI \ion{Ca}{II} K annular brightening cannot be reliably trusted, there is a general correspondence between its spatial location and the enhanced intensities observed in the ViSP raster images in the wings of $\lambda8542$. Coincidentally, raster images in the blue (e.g., at $-70\rm\,m\AA$ from line core) also show a slight dimming at the center of the brightening, giving the impression of a ring-like brightening. However, the significant temporal evolution of the structure suggested by the VBI images likely distorts the emissions in the raster due to the motion of the ViSP slit.

\subsubsection{Magnetic Field and Temperature Stratifications}

The $B_{\rm LOS}$ map in the photosphere clearly shows the small minority (negative) magnetic polarity patch, unlike the co-temporal HMI magnetogram of the same region (cf. Fig.~\ref{fig:timeseries}). The H$\beta$ brightening is located between the two opposite polarities, while the \ion{Ca}{II} K brightening is spatially offset in the limb-ward direction, but it partly overlaps with the EB. The ring structure is approximately centered on the negative polarity and traces a semi-circular polarity inversion line (PIL). While the magnitude of the $B_{\rm LOS}$ component of the minority negative polarity patch reaches up to $850$\,G at $\log\tau$\,$=$\,$-2$, it practically vanishes in higher layers, with the magnetograms showing an essentially unipolar field for optical depths lower than $\log\tau$\,$\sim$\,$-4$. 
However, the polarimetric sensitivity of these ViSP observations is insufficient to reliably determine the chromospheric vector field at a spatial resolution of 0.1\asec, with the denoised magnetic field maps looking quite patchy. However, high-resolution is clearly needed to resolve the fine-scale structures of the event, as highlighted by the VBI image sequences.

\begin{figure*}[ht]
    \centering
    \includegraphics[width=\linewidth]{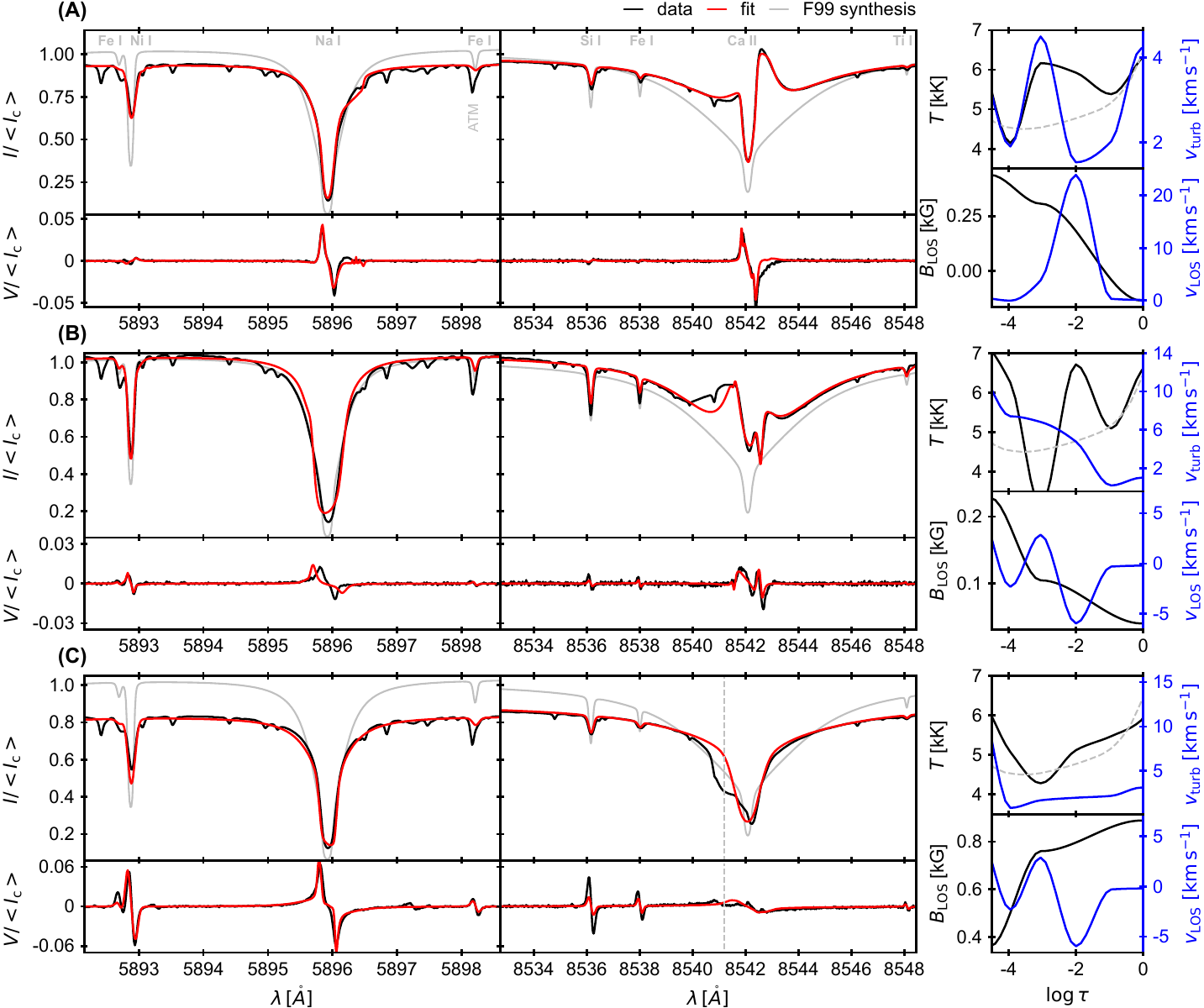}
    \caption{Spectra and best models at different locations. The three sets of panels show the observed ViSP spectra and best-fit models (left panels), as well as the corresponding model parameters as a function of logarithmic continuum optical depth (right panels)) at the three locations designated (A), (B), and (C) in Fig.~\ref{fig:inv}.  The gray dashed line in (C) indicates the position of the misfitted component at $-31\,\rm km\,s^{-1}$ relative to the nominal line center. The gray solid lines in the left panels show the synthetic spectra computed from the F99 model, whose temperature profile is shown on the right panels (dashed line) for comparison with the targets' profiles.}
    \label{fig:fits}
\end{figure*}

We find significant temperature enhancements in the photosphere around $\log\tau$\,$=$\,$-2$, coinciding with broadband intensity enhancements in the VBI filters. Interestingly, we do find a ring-like structure in the inverted temperatures at $\log\tau=-2$, as highlighted by the inset panel.
The temperature enhancements range from $\sim$\,100$-$1,000\,K around location (A) to just over 2,000\,K in the bright kernel at the top (e.g., location B) relative to the QS model, shown here for comparison.
Temperatures also increase at the center of the ring structure in somewhat higher layers around $\log\tau$\,$=$\,$-3$, which can be interpreted as heating occurring in a dome-like structure. However, the inversion maps are noisy at this depth due to the weak sensitivity of all diagnostics. Lower optical depths layers in the chromosphere ($\log\tau$\,$\sim$\,$-4.5$) do not show significant temperature enhancements, suggesting that the energy release is quite deep-seated. Consequently, the brightenings observed in the broad-band \ion{Ca}{II} K and H$\beta$ images (Fig.~\ref{fig:timeseries2}) are likely dominated by emission from the line wings, originating in the upper photosphere and lower chromosphere. This is also evident from the $\lambda$8542 profiles at locations (A) and (B) displayed in Fig.~\ref{fig:fits}, which show strong wing emissions but relatively weaker core enhancements compared to the QS profile. The temperature stratifications at these locations show broad bumps at optical depths ($\sim$\,2 dex) in the photosphere. 

We also calculated the radiative energy losses in the brightenings from our inversion models following \citet[][]{2021A&A...647A.188D}, and found values reaching up to $\sim$\,$70\rm~kW\,m^{-2}$ in the low chromosphere, integrated between the $T_{\rm min}$ height and the formation height of $\lambda$8542 core, with an uncertainty on the order of 15\% \citep[][]{2024arXiv240815908D}. This value is several times higher than the canonical cooling rates in the low chromosphere in ARs \citep[$\sim$\,$10\rm\,kW\,m^{-2}$;][]{1977ARA&A..15..363W}. However, at some locations, the temperature enhancements begin in deeper layers around $\log \tau\sim-1$ (e.g., location A and B). If we integrate the cooling rates from these depths, the total losses could increase by up to an order of magnitude due to the higher densities in the photosphere. For a brightening size of 1\asec~and duration of 5\,min, the total radiative energy would be $\sim$\,$10^{27}$\,erg \citep[see also][]{2006ApJ...643.1325F}. Using the same size as a characteristic length, $L$, and the measured flux cancellation rate from HMI over the same duration (Section~\ref{sec:obs_stats}), the magnetic energy, $B\Phi L/(8\pi)$, is of the same order $\sim$\,$10^{27}$\,erg \citep[see also][]{2020A&A...644A.130C}.

\subsubsection{Enhanced Velocities and Microturbulence}

The LOS velocity maps at the same optical depths show that the lower part of the brightening show mixed downflows and upflows, while the compact bright kernel and bright jet-like structures are mostly up-flowing, forming a bidirectional flow pattern at $\log \tau\sim-2$. Inferred velocities at this depth reach supersonic values up to $\sim$\,$28\rm \,km\,s^{-1}$ at the location of the EB brightenings between the two opposite polarities, and up to $\sim$\,$-18\rm \,km\,s^{-1}$ in the upper structures. Location (A) is one such cases, showing velocities peaking just over $\sim$\,$20\rm \,km\,s^{-1}$ at $\log\tau$\,$=$\,$-2$ (Fig.~\ref{fig:fits}). These velocity magnitudes are on the order of the Alfvén speed at that optical depth. The chromospheric velocities (lower panel in Fig.~\ref{fig:inv}) do not show any coherent structures associated to the brightenings other than an extended downflow patch in the upper part of the FOV. However, the inferred velocity maps do not capture the complete picture. 

At location (C), for example, the inversions cannot fit the $\lambda$8542 profiles associated to the dark structures visible in line wings and core. We observe a depression in the blue wing of $\lambda8542$ at about $\sim$\,$-31$\,$\rm km\,s^{-1}$ (vertical dashed line) relative to nominal line center, while the $\lambda5896$ line lacks that feature (Fig.~\ref{fig:fits}). We tentatively interpret this as part of a surge outflowing from the upper brightening. Surges are relatively cool outflows sometimes observed in conjunction with EBs and UVBs, driven by the magnetic reconnection process \citep[e.g.,][]{2011ApJ...736...71W,2021A&A...655A..28N}. Similar features have also been reproduced in simulations \citep[][]{2019A&A...626A..33H}. This suggests fast-moving, cool gas at higher-than-average altitudes, which cannot be modeled under a plane-parallel atmosphere in hydrostatic equilibrium. Other dark features in the core of $\lambda8542$ are visible in the bottom left panel in Fig.~\ref{fig:inv}; however, these are not associated with the brightenings but rather with the background chromospheric canopy. Notably, they do not show corresponding absorbing features in the blue line wing unlike the surge (center left in Fig.~\ref{fig:inv}).

Microturbulence is also greatly enhanced in the brightening, particularly around $\log \tau$\,$=$\,$-3$, showing a mean(max) value of $\sim$\,$5(14)\rm\,km\,s^{-1}$ while the FOV average is $\sim$\,$0.4\rm\,km\,s^{-1}$. Interestingly, this enhancement occurs in higher layers than the temperature enhancements, as exemplified by the profiles at locations (A) and (B) in Fig.~\ref{fig:fits}. Although we find both increased Doppler shifts and microturbulence in the brightenings, there is no correlation between their magnitudes at any optical depth.

\subsubsection{Challenges in Line Fitting}

The strongest ViSP lines were generally very well fitted in the quiet areas of the FOV, where simple absorption profiles make further discussion unnecessary.  
The examples displayed in Fig.~\ref{fig:fits} also show generally good fits to the $\lambda$5896 and $\lambda$8542 line profiles in the brightenings. Locations showing strong Doppler shifted emission or absorption in $\lambda$8542 were more challenging to fit even if we increased the number of nodes, causing some incompatibilities in the $\lambda$5896 modeling, specifically forcing the sodium line to be brighter than observed (e.g., location B). 
In addition, several weaker photospheric lines, which were not included in the inversion process, show varying degrees of reproducibility. For example, the \ion{Ni}{I} 5893\,\AA~is relatively well reproduced in both intensity and circular polarization, whereas the \ion{Fe}{I} line, blended in its blue wing, is clearly too deep in the observations compared to the model predictions. The polarization signals in the \ion{Si}{I} and \ion{Fe}{I} in the blue wing of $\lambda$8542 are also systematically underestimated across the FOV. While these discrepancies do not impact the inversion results presented here, further investigation is required to identify the sources of these systematic errors so that these additional spectral diagnostics can also be utilized in the future.

\subsection{Extrapolated Magnetic Configuration}
\label{sec:extrapol}

\begin{figure*}
    \centering
    \includegraphics[width=\linewidth]{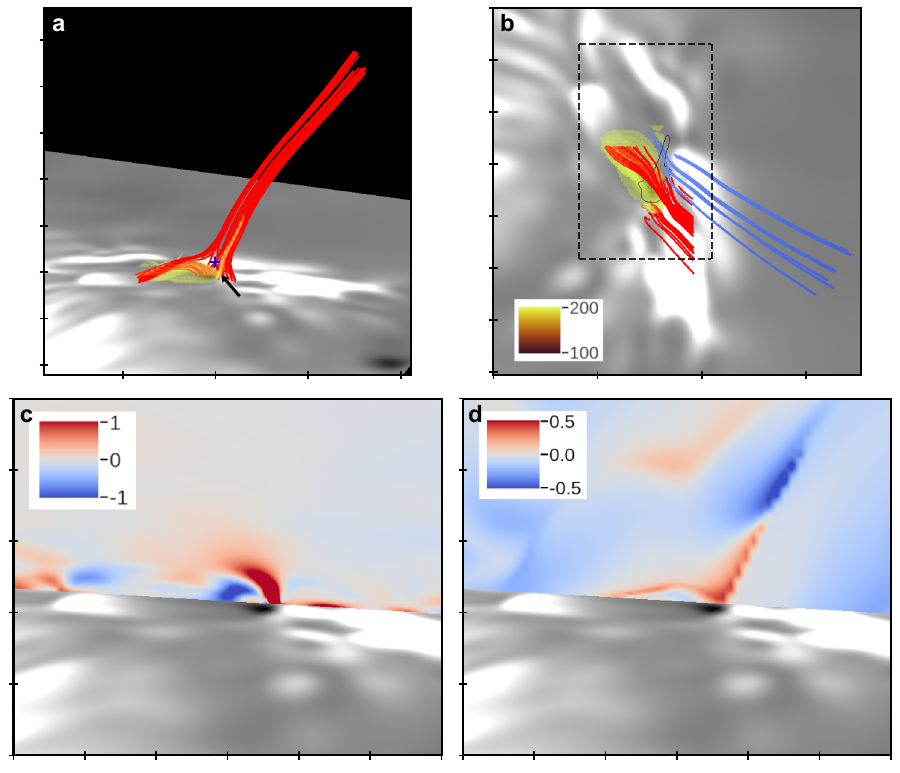}
    \caption{Magnetic field extrapolation  with the NLFFF method. (a): Side view of the ROI showing the traced field lines (red), and a volume rendering of the squashing factor (yellow); the blue cross shows the location of the magnetic null point, located $\sim$\,500 km above the magnetogram displayed at the lower boundary (capped at $\pm$\,700\,G.). (b): Top view of the ROI, including blue field lines traced from the footpoints of the dark features observed in the wing of $\lambda$8542 (cf. Fig.~\ref{fig:inv}); the solid black contours and dashed box mark the same contours and FOV as displayed in Fig.~\ref{fig:inv}. (c): Vertical slice of the field curvature. (d): Vertical slice of the twist number. All colormaps are capped for display purposes.}
    \label{fig:extrapol}
\end{figure*}

Figure~\ref{fig:extrapol} shows the extrapolated magnetic field lines from the photospheric vector magnetograms, including a volume rendering of the squashing factor (yellow shade, (a)$-$(b)). We also show a vertical slice of the field curvature (c) and twist number (d).
The visualization was created using \texttt{VAPOR} \citep{2019Atmos..10..488L}. The red field lines outline a low-lying dome structure connected to a spine, resembling a fan-spine or "anemone jet" configuration \citep[e.g.,][]{1995Natur.375...42Y,science.1146708}. 
This dome structure is also evident in the curvature map (c), showing a transition from a compact region with negative (concave) curvature to positive curvature above it. From the variation of the sign of the field curvature with height ($\hat{B} \cdot \nabla \hat{B}_z$), we determine the dome height to extend up to $\sim$700\,km above the photospheric layer, which is within the expected formation height range of $\lambda$8542. The extrapolated field strength is on the order of 300\,G or less at that height, consistent with the inferred magnetic field from the NLTE inversions around $\log \tau$\,$\sim$\,$-4.5$ within the uncertainties. This compact dome is also consistent with the small negative polarity field patch vanishing from the inversion maps quickly with height (cf. Fig.~\ref{fig:inv}).

A magnetic null point was identified $\sim$500\,km above the extrapolation’s lower boundary, located between the two opposite polarities on one side of the dome and connecting to the spine. The projection of this null-point on the $z=0$ plane coincides with the EB brightenings; however, it alone cannot fully explain the event's complex morphology. Specifically, the bright fibril-like (and jet-like) structures oriented toward northeast (and east) in the \ion{Ca}{II} K images (e.g., Fig.~\ref{fig:timeseries2}, 20:38\,UT), are positioned a few arcseconds away from the null point and are not aligned along the main spine but rather at an angle to it. 
The dark blue-shifted structures observed in $\lambda$8542 (cf. Fig.~\ref{fig:inv}) are also shifted by $\sim$\,1$-$2\asec~to the north relative to the main spine. The blue-colored field lines in this region indicate that the inferred topology roughly follows the orientation of the absorbing features. It is plausible that reconnection occurs at multiple locations across the structure that are not necessarily null points.

The squashing factor, $Q$, also shows a low-lying dome-like structure. High $Q$ values (a few $\sim$\,$10^2$) are concentrated along the boundaries of the dome, indicating sharp gradients in magnetic field connectivity. These regions are quasi-separatrix layers (QSLs), where magnetic reconnection can occur \citep[e.g.,][]{1995JGR...10023443P,2002JGRA..107.1164T}.
The QSL encloses the fan field lines, while the spine extends through the dome’s top. The current density is also enhanced around the dome in a semi-circular region below 1\,Mm in height tracing the PIL (not displayed). However, the extent of the Q surface is significantly larger than the \ion{Ca}{II} K brightenings, extending northwest into regions where no enhanced emission is observed.
The highest $Q$ values ($\sim$\,$10^3-10^4$) appear in an extended, narrow sheet (black arrow) that stretches 1.4 Mm upward from the boundary along the spine between the two opposite polarities. This region also coincides with the location of the H$\beta$ brightenings. 

The twist number map shows localized enhancements between the two polarities, possibly due to shearing or rotational motions of the dome/spine footpoints. This may contribute to magnetic stress buildup, leading to high $Q$ values. While the twist numbers are generally small ($|\mathcal{T}_w|$\,$\lesssim0.8$), a clear transition from negative to positive twist occurs with height at the dome and along the spine.

We note that the ViSP only captured a narrow time interval of a series of recurrent brightenings at the same location, as shown in Fig.~\ref{fig:timeseries} and Fig.~\ref{fig:timeseries2}. These successive brightenings may have driven the magnetic structure into a lower-energy state at this instance. Note that the extrapolation requires a trade-off between the observed photospheric magnetic field and the force-free assumption, resulting in an intrinsically smoother modeled boundary condition (Fig.~\ref{fig:extrapol}). Additionally, the extrapolated field may not fully capture certain aspects of the field configuration, given the low cadence of the ViSP raster relative to the rapid timescale of the event.

\section{Discussion and Conclusions}

In this paper, we present high-resolution observations of a series of brightenings in an active region near the solar disk center. We interpret them as the result of a low-altitude, small-scale magnetic reconnection event occurring as part of a long-lived flux cancellation process. We observe a steady decrease in magnetic flux observed in HMI magnetograms over an hour and recurrent 1700\,\AA~continuum enhancements at the same location. Although the ViSP raster captured only a brief, few-minute snapshot of this process, it enabled the detection of strong Doppler shifts and increased microturbulence in spectral lines probing the upper photosphere and low chromosphere. The velocity magnitudes are comparable to the local Alfvén speed, consistent with reconnection-driven flows \citep[e.g.,][]{2022LRSP...19....1P}. 

The broadband VBI images reveal EB-like flickering emissions and compact blobs in H$\beta$, whereas \ion{Ca}{II}\,K shows a more extended bright annular structure and hot jet-like outflows. The bright ring was found to be an artifact of the image calibration caused by the overexposure in that filter. Work is underway to improve the robustness of the image reconstruction. Interestingly, the inversion results based on the ViSP data do show a compact ($\sim 1^{\prime\prime}$) annular region with enhanced temperatures at $\log \tau=-2$, suggesting that the VBI emission may partly come from a circular region around the brightening. Temperatures are also enhanced at the brightening center at slightly lower optical depths. The VBI image overexposure at the brightnening center could then imply strong radiative cooling from slightly higher layers in the low chromosphere at the center of the brightening relative to its surroundings. However, due to the limited image quality in the \ion{Ca}{II}\,K filter, we are unable to reliably cross-validate the results of the ViSP data analysis with the imaging observations in that filter.

Previous observations have shown bright substructures in magnetic reconnection events, suggesting plasmoid-mediated intermittent reconnection \citep[e.g.,][]{2015ApJ...813...86I,2017ApJ...851L...6R, 2021A&A...647A.188D,2025arXiv250110246L}, or energy deposition in small-scale magnetic loops \citep[][]{2018A&A...617A.128S}. In the \ion{H}{$\beta$} image sequence, we observe structures that align with the plasmoid interpretation. However, these structures may not necessarily be magnetic islands; they could instead be hot, over-dense regions driven by shocks and/or turbulence, emitting strongly in H$\beta$. Discriminating between these scenarios requires higher-cadence ($<$30\,s) and higher-sensitivity magnetometry, which is a technical challenge for slit-based spectropolarimeters like the ViSP. 

These observations suggest a more complex magnetic topology than that of classical EBs \citep[e.g.,][]{2013JPhCS.440a2007R} resembling cases in \citet[][]{2024arXiv241203211B}. In that study, some QS EBs were found to have UV emission counterparts originating from different parts of fan-spine structures, as revealed by magnetic field extrapolations. Similarly, some AR EBs also have UVB counterparts in IRIS data, which can be spatially offset or display different characteristics \citep[e.g.,][]{2015ApJ...812...11V,2016ApJ...824...96T,2019ApJ...875L..30C,2020A&A...633A..58O}. However, at a spatial resolution of $\sim$\,0.33-0.4\asec, IRIS brightenings typically appear compact (or dot-like), elongated, or surge-like \citep[e.g.,][]{2019ApJ...887...56T}. 
Follow-up high-resolution observations with the VBI are needed to investigate potential substructure in these events to provide critical clues about the physical mechanisms at play, such as whether energy is being released in a bursty or more continuous fashion, or whether energy deposition by nonthermal particles is significant \citep[e.g.,][]{2023ApJ...956...85T}. Likewise, co-observations with IRIS are needed to explore potential UV spectral signatures. Regardless of the event classification, these observations provide further evidence of the various ways in which magnetic reconnection can occur in the solar atmosphere \citep[reviewed in ][]{2017JPlPh..83a5301J,2022LRSP...19....1P}.

This interpretation is further supported by a multi-resolution NLFFF extrapolation based on a combination of HMI and ViSP magnetograms. The magnetic topology resembles a fan-spine configuration with a null point $\sim$500 km above the photosphere. Coincidentally, our height estimate matches that provided by \citet[][]{2017A&A...605A..49C} where a \ion{Si}{IV}\,1400\,\AA~UVB observed by IRIS coincided with a null-point location in a fan-spine structure. However, in our case, brightenings extend beyond the null, suggesting a different origin. 
We also find enhanced squashing factors within a low-lying, dome-shaped structure surrounding the negative polarity patch, extending up to $\sim$\,0.7\,Mm in height, as well as within an extended thin sheet between the two opposite polarities, reaching up to $\sim$\,1.4\,Mm. These regions define QSLs, where three-dimensional reconnection can occur without null points through continuous magnetic field line slippage \citep[e.g.,][]{1995JGR...10023443P,1996JGR...101.7631D}. This heated magnetic dome may tentatively explain the temperature stratification and increased microturbulence inferred from the inversions. 

The continuous inflow of magnetic bright elements at relatively high speeds ($\sim$4$-$6\,$\rm km\,s^{-1}$) to the region, as observed in the G-band image sequences when the ring appears (compared to their absence earlier in the time series), suggests that they may have contributed to significant magnetic field gradients, including field twist. This likely stressed the magnetic structure and led to enhanced squashing factors. 
Therefore, reconnection at QSLs plausibly explains the emissions \citep[see also][]{2004ApJ...614.1099P, 2018ApJ...854..174T}, with physical variations of the QSL due to magnetic reconnection and force rebalancing and/or motions by the photospheric bright points potentially explaining the significant temporal evolution of the event  hinted by the \ion{Ca}{II} K images. We note that this process unfolded in a region already undergoing flux cancellation, as shown by the persistent EB-like H$\beta$ brightenings between the opposite magnetic polarities before the \ion{Ca}{II} K brightenings became visible.

Regarding the inferred temperatures in the EB, both the $\lambda$5896 and $\lambda$8542 lines can be well reproduced without requiring a highly localized temperature hump ($\lesssim$\,20\,km) in the upper photosphere \citep[][]{2019ApJ...871..125S}. These lines can be well fitted with relatively broad temperature peaks ($\Delta T$\,$\sim$\,1\,kK) over a range of optical depths corresponding to a couple hundred kilometers assuming hydrostatic equilibrium. 
However, the high downflow speeds inferred at the EB locations ($\sim$\,20\,$\rm km\,s^{-1}$) may slightly increase the effective gravity and decrease the pressure scale height. 
Notably, the event remains invisible in sodium line intensities. However, according to our models, this does not necessarily imply temperatures high enough to fully ionize the neutral species in those layers \citep[$\gtrsim$\,10\,kK,][]{2015ApJ...808..133R,2016A&A...590A.124R}. In contrast, observations in \ion{He}{I} 10830\,\AA~show that significantly higher temperatures ($\gtrsim$\,20\,kK) than the ones inferred here occur in some EBs deep in the atmosphere \citep[][]{2017A&A...598A..33L,2025arXiv250110246L}, beyond the sensitivity range of the \ion{Na}{I} and \ion{Ca}{II} lines. 

To advance our understanding of the energetics and dynamics of these events, future efforts should focus on acquiring multi-height, multi-wavelength datasets with high temporal resolution. This will be necessary to improve reconstructions of the magnetic field topology, particularly through multi-layer extrapolations \citep{Jarolim2024multi}.

\begin{acknowledgments}
We thank Friedrich Wöger, Alisdair Davey, and Alexandra Tritschler for the insightful discussions regarding VBI image processing.
The research reported herein is based in part on data collected with the Daniel K. Inouye Solar Telescope (DKIST), a facility of the National Solar Observatory (NSO). DKIST is located on land of spiritual and cultural significance to Native Hawaiian people. The use of this important site to further scientific knowledge is done so with appreciation and respect. The NSO is operated by the Association of Universities for Research in Astronomy, Inc., under cooperative agreement with the National Science Foundation. 
This research was supported by the National Science Foundation REU program, Award \#1950911. 
This work utilized the Blanca condo computing resource and the Alpine high performance computing resource at the University of Colorado Boulder. Blanca is jointly funded by computing users and the University of Colorado Boulder. Alpine is jointly funded by the University of Colorado Boulder, the University of Colorado Anschutz, and Colorado State University. This project has received funding from the Swedish Research Council (2021-05613) and the Swedish National Space Agency (2021-00116). We acknowledge resources provided by the National Academic Infrastructure for Supercomputing in Sweden (projects NAISS 2023/1-15 and NAISS 2024/1-14) at the PDC Centre for High Performance Computing (PDC-HPC) at the Royal Institute of Technology in Stockholm.
\end{acknowledgments}

%

\vspace{5mm}
\facilities{SDO(AIA \& HMI), DKIST(ViSP \& VBI)}


\software{ \texttt{Astropy} \citep{2013A&A...558A..33A,2018AJ....156..123A},
\texttt{Sunpy} \citep{sunpy_community2020}, 
\texttt{pyMilne} \citep{2019A&A...631A.153D},
\texttt{STiC} \citep{2019A&A...623A..74D},
\texttt{NF2} \citep{jarolim2023nf2},
\texttt{FastQSL} \citep{Zhang2022fastqsl}, 
\texttt{VAPOR} \citep{2019Atmos..10..488L}
}

\bibliographystyle{aasjournal}



\end{document}